\newcommand{\beq}{\begin{eqnarray}}
\newcommand{\eeq}{\end{eqnarray}}

\documentclass[preprintnumbers,amsmath,amssymb,pra]{revtex4}

\usepackage{graphicx}

\begin{document}

\title{The twin paradox in a cosmological context}

\author{\O yvind Gr\o n}
\email{Oyvind.Gron@iu.hio.no}
\author{Simen Braeck}
\affiliation{Oslo University College, Faculty of Engineering, P.O. Box 4 St. Olavs Plass, N-0130 Oslo, Norway}

\begin{abstract}
Recently Abramowicz and Bajtlik [ArXiv: 0905.2428 (2009)] have studied the twin paradox in Schwarzschild
spacetime. Considering circular motion they showed that the twin with a non-vanishing 4-acceleration is older
than his brother at the reunion and argued that in spaces that are asymptotically Minkowskian there exists an
absolute standard of rest determining which twin is oldest at the reunion. Here we show that with vertical motion
in Schwarzschild spacetime the result is opposite: The twin with a non-vanishing 4-acceleration is younger.
We also deduce the existence of a new relativistic time
effect, that there is either a time dilation or an increased rate of time associated with a clock moving in a
rotating frame. This is in fact a first order effect in the velocity of the clock, and must be taken into account
if the situation presented by Abramowicz and Bajtlik is described from the rotating rest frame of one of the twins.
Our analysis shows that this effect has a Machian character since the rotating state of a frame depends upon the
motion of the cosmic matter due to the inertial dragging it causes. We argue that a consistent formulation
and resolution of the twin paradox makes use of the general principle of relativity and requires the
introduction of an extended model of the Minkowski spacetime. In the extended model Minkowski spacetime
is supplied with a cosmic shell of matter with radius equal to its own Schwarzschild radius, so that
there is perfect inertial dragging inside the shell.

\end{abstract}

\maketitle

\section{Introduction\label{sec:intro}}

For nearly a hundred years the twin paradox has been a source of inspiration and wonder for people
learning and teaching the theory of relativity. Usually one twin, A, stays at rest in the flat Minkowski
spacetime and the other, B, travels away and then comes back again. Then B is younger than A when they meet again.
However from the point of view of B, and according to the principle of relativity, B could regard himself
as at rest and A as the traveler. Then A should be younger than B when they meet again.

The standard resolution is that the twin who has accelerated mostly while they were away from each other is
the younger one when they meet
again~\cite{Marder,EriksenGron,DebsRedhead,Nicolic,Iorio,GronHervik,LichteneggerIorio}.
In other words \emph{the accelerated twin is younger}. The confusion that leads to the apparent paradox arises
from an incorrect application of the principle of special relativity, which can only be applied to inertial
observers. However, in two very interesting articles M. A. Abramowicz and
S. Bajtlik~\cite{AbramowiczBajtlik,AbramowiczBajtlikphoton} have presented some versions of the twin
paradox which point to a different result. In the last one they state in the heading that "the accelerated
twin is older". Here A is at rest in the Schwarzschild space outside a non-rotating
star and B moves in a circular orbit. A short calculation shows that A has aged more than B when they meet
after B has moved around the star. B is moving freely and has vanishing 4-acceleration, while A is acted
upon by a force keeping him at rest, and hence has a non-vanishing 4-acceleration. This is the reason for
the statement "the accelerated twin is older" in their heading.

It should be noted, however, that A has no 3-acceleration while B has a non-vanishing centripetal acceleration.
Hence, like in Minkowski spacetime, the twin with vanishing 3-acceleration is older. In general 3-acceleration
and 4-acceleration have different physical meanings and different properties. The three-acceleration is
relative and can be transformed away by going into the co-moving rest frame of the object. The 4-acceleration
of an object is absolute and cannot be transformed away. It is due to non-gravitational forces and vanishes
only for a freely falling object.

In the "standard resolution" of the twin paradox mentioned above the description of the twins are given with
reference to an inertial reference frame in Minkowski spacetime, and the "acceleration" is usually meant to
be a 3-acceleration. However, in this case there is a degeneracy between the 3-acceleration and the
4-acceleration. For twins in flat spacetime the invariant statement would be: The twin with a non-vanishing
4-acceleration is younger. It is this statement that Abramowicz and Bajtlik have shown is not generally true.

They gave some suggestions about what could be true. In this connection they introduced what they called an
"absolute standard of rest" in the Schwarzschild spacetime, writing: "The absolute standard of rest in the
Schwarzschild spacetime is given by the local Killing symmetry, $\partial_t=0$, the non-zero curvature, and the
global condition of the asymptotical flatness (the starry sky above)".

For the situation referred to above they deduced that the ageing of A and B is related by
\beq
\frac{\tau_A}{\tau_B}=\sqrt{\frac{1-v_A^2/c^2}{1-v_B^2/c^2}}
\label{eq:Abramowiczratio}
\eeq
with $v_A=0$, and write that this formula "reveals that the ratio of proper times measured by the twins between
their two consecutive reunions depends only on their orbital velocities $v_A$ and $v_B$, measured with respect to
the absolute global standard of rest."

Considering some different situations they concluded: "In all situations in which the absolute motion
may be defined in terms of some invariant \emph{global} properties of spacetime, the twin who moves faster with
respect to the global standard of rest is younger at the reunion, \emph{irrespectively to twins' accelerations}."
Then they asked: "Could the notion "\emph{the twin who moves faster, is younger at the reunion}" be somehow extended
to the classical version of the paradox in the Minkowski spacetime, for example by referring to the starry
sky above the twins?". They left this question unanswered.

That the resolution of the twin paradox is somehow connected with the "starry sky" was noted by Einstein already
more than seventy years ago~\cite{Einstein}. Neglecting localized mass distributions and writing about the twin paradox
from the point of view of a non-inertial twin (the one who must be acted upon by non-gravitational forces) he
says that all the stars in the world are accelerated relative to this twin, and they then induce a gravitational
field analogous to the electrical field induced inside an accelerated electrically charged
shell~\cite{GronEriksen}. More recently B. R. Holstein and A. R. Swift~\cite{HolsteinSwift} wrote: "The
Earth-bound twin is at rest relative to the Universe, while his brother accelerates relative to the Universe.
In the frame of the traveling twin, his brother and the entire Universe moves away and returns. This
accelerating Universe generates a gravitational field which slows his clocks."

They further wrote: "It is difficult for us to see how the two observers could detect the asymmetry in their
motion, without observations either of the central mass or of the fixed stars." One may ask: What about
accelerometers? Note that Holstein and Swift ask for a means of detecting acceleration if the central mass
and the fixed stars are removed, i.e. in a universe that is empty except for the twins. In other words: How
can one determine acceleration in a globally empty Minkowski spacetime.

Our point of view is that this cannot be done within a theory where absolute motion does not exist. In the
globally empty Minkowski spacetime the choice of an inertial reference frame has an ad hoc character. We need
an extended model of Minkowski spacetime to remove this incompleteness of the theory. This is introduced and
further discussed in section VII below.

In the present article we shall try to give a general answer to the question of Abramowicz and Bajtlik cited
above. We approach the question
by considering some versions of the twin paradox in a field of gravity, first in a uniformly accelerated
reference frame in flat spacetime, then in a Friedmann-Robertson-Walker (FRW) universe model and finally
by considering two twins in the Schwarzschild and Kerr spacetimes.

We point out the significance of perfect inertial dragging~\cite{Gron} for determining the cosmic inertial
frames and formulate a supposition which is  a first step towards the answer to Abramowicz's and Bajtlik's question.
However, a new version of the twin paradox in which one clock is thrown vertically upwards in the Schwarzschild
space and then falls down again to a clock remaining at rest, shows that our supposition cannot be the
final answer to Abramowicz's and Bajtlik's question. Our final answer is: The notion that "the twin who moves faster
is younger at the reunion" can be extended to the classical version of the twin paradox in the Minkowski
spacetime as described in an inertial (non-rotating) reference frame. But it is not valid in general. The rate
of ageing of a twin depends upon three effects; the velocity dependent time dilation, the gravitational time
dilation and, as we shall see below, a cosmic time effect. Hence there is no general recipe, but only a mix of
strategies to age slowly. If two twins compare their ages at two events $P_1$ and $P_2$, the twin
who "on the average" has moved most slowly relative to the cosmic matter and been highest up in a gravitational
field, will be oldest at the second event $P_2$.

\section{The twin paradox in a uniformly accelerated reference frame\label{sec:uniformacc}}

The twin paradox has been treated in hundreds of articles. In addition to
Refs.~\cite{EriksenGron,DebsRedhead,Nicolic,Iorio,GronHervik,LichteneggerIorio} the twin paradox has been analyzed with
reference to uniformly accelerated reference frames in flat spacetime in Refs.~\cite{Boughn,Styer,GiannoniGron}.
In the present section we treat vertical motion in a uniformly accelerated reference frame.

Let the uniformly accelerated reference frame have coordinates $(t,x,y,z)$. We consider two twins A and B.
Twin A stays at rest in the uniformly accelerated reference frame and has a constant rest acceleration $g$
in the $x$-direction.
The traveling twin, B, starts from the origin $x=0$ with an initial coordinate 3-velocity
$v_0=(dx/dt)_{t=0}$ in the positive x-direction and moves freely along the $x$ axis. Thus, according to twin A,
twin B is shot upwards and is then freely falling in a uniform gravitational field.

In the uniformly accelerated reference frame the line element is given by
\beq
ds^2 = -\left(1+\frac{gx}{c^2}\right)^2c^2dt^2 + dx^2 + dy^2 + dz^2 \,,
\label{eq:uniaccle}
\eeq
and the Lagrangian of B can be written
\beq
L = -\frac{1}{2}\left(1+\frac{gx}{c^2}\right)^2c^2{\dot{t}}^2 + \frac{1}{2}{\dot{x}}^2\,,
\label{eq:uniLagrangian}
\eeq
where the dots denote derivatives with respect to the proper time $\tau$ of B. Since the Lagrangian is independent
of $t$, the momentum conjugate to the time-coordinate
\beq
p_t = \frac{\partial L}{\partial\dot{t}}=-\left(1+\frac{gx}{c^2}\right)^2c^2\dot{t}
\label{eq:uniconstantofmotion}
\eeq
is a constant of motion. Utilizing Eq.~(\ref{eq:uniconstantofmotion}), the 4-velocity identity
$\vec{u}\cdot\vec{u}=-c^2$ gives
\beq
p_t^2-\left(1+\frac{gx}{c^2}\right)^2c^4 = \left(1+\frac{gx}{c^2}\right)^2c^2{\dot{x}}^2\,.
\label{eq:univelidentity}
\eeq
$p_t$ may now be determined by observing that, at the instant when twin B turns around and reaches the
highest point $x=h$ of his path, $\dot{x}=0$ and Eq.~(\ref{eq:univelidentity}) then yields
\beq
p_t=-\left(1+\frac{gh}{c^2}\right) c^2\,.
\label{eq:uniconstantfixed}
\eeq
By substituting the expression~(\ref{eq:uniconstantfixed}) for $p_t$ in Eq.~(\ref{eq:univelidentity}),
and noting that the travel time down is the same as the travel time upwards, the proper-time interval of
twin B during his travel becomes
\beq
\tau = 2\int_{0}^{h}\frac{1+\frac{gx}{c^2}}{\sqrt{\left(1+\frac{gh}{c^2}\right)^2c^2
-\left(1+\frac{gx}{c^2}\right)^2c^2}}\,dx \,.
\label{eq:uniproptimeint}
\eeq
The integral gives
\beq
\tau = \frac{2c}{g}\sqrt{\left(1+\frac{gh}{c^2}\right)^2-1}\,.
\label{eq:uniproptime}
\eeq

To calculate the corresponding time interval as measured by twin A's clock, we combine
Eqs.~(\ref{eq:uniconstantofmotion}) and (\ref{eq:uniconstantfixed}) to obtain
\beq
\frac{dt}{d\tau} = \frac{1+\frac{gh}{c^2}}{\left(1+\frac{gx}{c^2}\right)^2}\,.
\label{eq:unidifftimes}
\eeq
Using that $dt/d\tau=(dt/dx)(dx/d\tau)$ and substituting for $dx/d\tau$ from
Eq.~(\ref{eq:uniproptimeint}), we find
\beq
t = 2\int_{0}^{h}\frac{1+\frac{gh}{c^2}}{\left(1+\frac{gx}{c^2}\right)
\sqrt{\left(1+\frac{gh}{c^2}\right)^2c^2-\left(1+\frac{gx}{c^2}\right)^2c^2}}\,dx\,. \nonumber \\
\label{eq:unicoordtimeint}
\eeq
Evaluating this integral, we get
\beq
t = \frac{2c}{g}\ln\left(1+\frac{gh}{c^2}+\sqrt{\left(1+\frac{gh}{c^2}\right)^2-1}\right)\,.
%t = \frac{2c}{g}\operatorname{arccosh}\left(1+\frac{gh}{c^2}\right)\,.
\label{eq:unicoordtime}
\eeq
From Eqs.~(\ref{eq:uniproptime}) and (\ref{eq:unicoordtime}) it now follows that
\beq
\frac{gt}{2c} = \ln\left(\frac{g\tau}{2c}+\sqrt{1+\left(\frac{g\tau}{2c}\right)^2}\right)
= \operatorname{arcsinh}\left(\frac{g\tau}{2c}\right)\,.
\label{eq:unireltimeproper}
\eeq
This equation may be rewritten as
\beq
\frac{g\tau}{2c}=\sinh\left(\frac{gt}{2c}\right)\,.
\label{eq:taurelt}
\eeq
Since $\sinh(gt/2c)>gt/2c$, it follows that $\tau>t$. In other words,
the traveling twin (twin B) is older than the twin who stays at rest (twin A) at the reunion.
This equation has been deduced in a different way by E.~Minguzzi~\cite{Minguzzi}.

The situation described here is similar to the situation discussed by Abramowicz and Bajtlik. Twin A, who is at
rest and ages by the time $t$, has vanishing coordinate 3-acceleration but is not freely falling and
therefore has a non-vanishing 4-acceleration. On the other hand, twin B, who is traveling and ages
by the time $\tau$, has a non-vanishing coordinate 3-acceleration but is freely falling and therefore has
a vanishing 4-acceleration. Yet, in this case, the twin who has a non-vanishing coordinate 3-acceleration,
vanishing 4-acceleration and moves faster is \emph{older} at the reunion, in contrast to what was found
in the example presented by Abramowicz and Bajtlik. We conclude, therefore, that 4-acceleration,
coordinate 3-acceleration and coordinate velocity cannot be decisive factors in determining which twin
becomes the older.

\section{The twin paradox in the FRW universe models\label{sec:FRW}}

We now consider vertical motion in an expanding, homogeneous and isotropic
universe model. Let the cosmic reference frame, defined by a set of freely moving
particles, have comoving coordinates $(t,r,\theta,\phi)$. Then the spacetime geometry of this universe is
described by the line element (see e.g. Ref.~\cite{GronHervik})
\beq
ds^2=-c^2dt^2
 +a^2(t)\left[\frac{dr^2}{1-kr^2}+r^2\left(d\theta^2+\sin^2\theta d\phi^2\right)\right]\,,
\label{eq:FRWmetric}
\eeq
where $a(t)$ is the scale factor. We now assume that twin A stays at rest
at the coordinate position $r=0$ while twin B performs vertical motion along the $r$-direction. At time
$t=0$ twin B starts from the position $r=0$ with a coordinate velocity $v_0=\left(dr/dt\right)_{t=0}$
in the positive $r$-direction and moves with a negative acceleration $-g(t)$ for times $t>0$. The
coordinate velocity of twin B at an arbitrary point of time is $v(t)$.

The traveling time $t_1$ measured by twin A's clock at reunion is obtained from the condition
$r(t)=0$. The corresponding time interval measured by twin B's clock can be
obtained from Eq.~(\ref{eq:FRWmetric}), giving
\beq
\tau=\int_{0}^{t_1}\sqrt{1-\frac{a^2\left(t\right)v^2}{\left(1-kr^2\right)c^2}}\,dt\,.
\label{eq:FRWproptime}
\eeq
Because the integrand is less than unity for all times $t$ we see that $\tau<t_1$. Thus twin B, who travels
and has non-vanishing 3-acceleration, non-vanishing 4-acceleration and moves faster, is \emph{younger} at the
reunion. Accordingly, the twin paradox in an expanding universe has a solution analogous to the classical
twin paradox in Minkowski spacetime. Note, however, that whereas Minkowski spacetime is flat the spacetime
of the FRW universe is curved. Thus, in determining which twin becomes the older, the effect of spacetime
curvature does not seem to be of any relevance.

\section{The twin paradox with vertical motion in the Schwarzschild spacetime\label{sec:Schwarzschild}}

In Schwarzschild coordinates $(t,r,\theta,\phi)$, the Schwarzschild geometry is described by
the line element
\beq
ds^2 = -\left(1-\frac{R_S}{r}\right)c^2dt^2 + \left(1-\frac{R_S}{r}\right)^{-1}dr^2
 + \,r^2\left(d\theta^2+\sin^2\theta d\phi^2\right)\,,
\label{eq:Schwmetric}
\eeq
where $R_S$ denotes the Schwarzschild radius. We shall now assume that twin A stays at rest at a fixed radius
$r=R$ during twin B's travel. Twin B starts from the same radius $r=R$ with an initial coordinate
3-velocity $v_0=(dr/dt)_{t=0}$ in the positive $r$-direction and continues to move freely in the radial
direction. The Lagrangian is then
\beq
L = -\frac{c^2}{2}\left(1-\frac{R_s}{r}\right){\dot{t}}^2 + \frac{1}{2}\frac{{\dot{r}}^2}{1-\frac{R_s}{r}}\,,
\label{eq:SchwLag}
\eeq
and the constant momentum conjugate to the time-coordinate thus becomes
\beq
p_t = -\left(1-\frac{R_s}{r}\right)c^2\dot{t} \,.
\label{eq:Schwconstmotion}
\eeq
The four-velocity identity now yields
\beq
p_t^2-\left(1-\frac{R_s}{r}\right)c^4 = c^2{\dot{r}}^2\,.
\label{eq:Schwidentity}
\eeq
$p_t$ is determined by noting that, when twin B reaches the highest point $r=R+h$, $\dot{r}=0$ and hence
\beq
p_t = -\sqrt{1-\frac{R_s}{R+h}}\,c \,.
\label{eq:Schwconstmotionfixed}
\eeq
Twin B's proper traveling time at the reunion is now found by substituting the expression for $p_t$
in Eq.~(\ref{eq:Schwidentity}):
\beq
\tau = \frac{2}{c}\sqrt{\frac{R+h}{R_s}}\int_{R}^{R+h}\sqrt{\frac{r}{R+h-r}}\,dr\,.
\label{eq:Schwproptimeint}
\eeq
Here the factor 2 accounts for both the travel time away and the travel time back. Carrying out the
integration then gives
\beq
\tau = \frac{2}{c}\sqrt{\frac{R+h}{R_s}}\left(\sqrt{Rh}+\left(R+h\right)
\arccos\sqrt{\frac{R}{R+h}}\right)\,. \nonumber \\
\label{eq:Schwproptime}
\eeq

\begin{figure}[btp]
\begin{center}
\includegraphics[width=8.0cm]{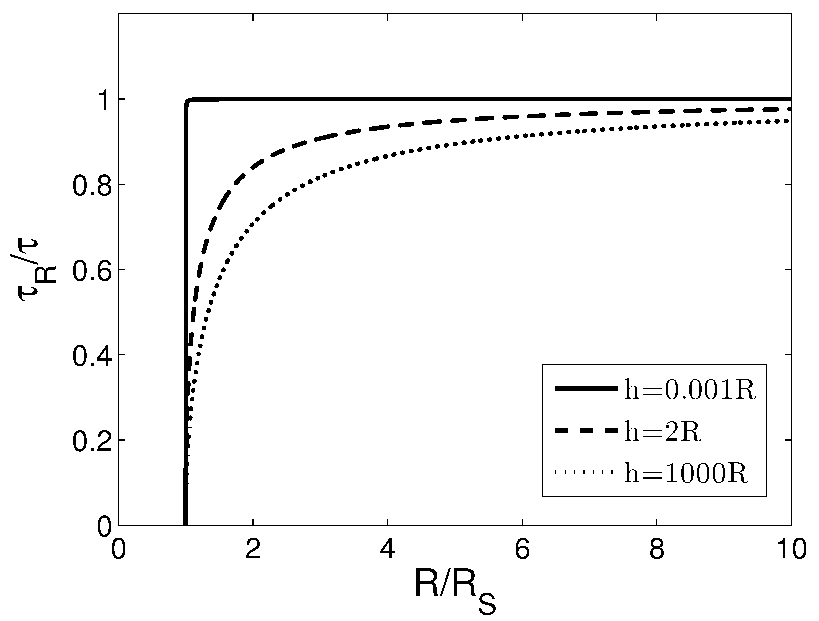}
\caption{The ratio of the traveling times measured by twin A's and B's clocks for different start
positions $R$ in Schwarzschild spacetime. Three curves are calculated corresponding to different choices of the
height $h$.}
\label{fig:Schwarztimeratios}
\end{center}
\end{figure}

In order to calculate the corresponding time interval $\tau_R$ shown by twin A's clock, we proceed to
combine Eqs.~(\ref{eq:Schwconstmotion}) and (\ref{eq:Schwconstmotionfixed}), yielding
\beq
\frac{dt}{d\tau} = \frac{\sqrt{1-\frac{R_s}{R+h}}}{1-\frac{R_s}{r}}\,.
\label{eq:Schwdifftimes}
\eeq
From Eq.~(\ref{eq:Schwdifftimes}), Eq.~(\ref{eq:Schwproptimeint}) and the relation $dt/d\tau=(dt/dr)(dr/d\tau)$
it now follows that
\beq
\frac{dt}{dr} = \frac{2}{c}\sqrt{\frac{R+h}{R_s}-1}\frac{r^{3/2}}{\left(r-R_s\right)\sqrt{R+h-r}}\,.
\label{eq:Schwdifftimerad}
\eeq
However, the coordinate time $t$ represents the time interval as measured by an observer at rest
infinitely far away from the central mass. From the line element it is seen that the relation
between coordinate time and the time interval shown on twin A's clock, which is at rest at the position $r=R$,
is $d\tau_R=\sqrt{1-R_s/R}\,dt$. Substituting this relation in Eq.~(\ref{eq:Schwdifftimerad}), we obtain
\beq
\tau_R = \frac{2}{c}\sqrt{\frac{\left(R-R_S\right)\left(R+h-R_S\right)}{RR_S}}
\int_{R}^{R+h}\frac{r^{3/2}}{\left(r-R_s\right)\sqrt{R+h-r}}\,dr\,.
\label{eq:Schwresttimeint}
\eeq
For the situation of interest here, $r>R_S$, and the integral in~(\ref{eq:Schwresttimeint}) then gives
\beq
\tau_R = \frac{2}{c}\sqrt{\frac{h\left(R-R_S\right)\left(R+h-R_S\right)}{R_S}}
 + \frac{4R_S+2R+2h}{c}
\sqrt{\frac{\left(R-R_S\right)\left(R+h-R_S\right)}{RR_S}}
\operatorname{arccot}\sqrt{\frac{R}{h}} \nonumber \\
 + \frac{4R_S}{c}\sqrt{\frac{R-R_S}{R}}\operatorname{arcoth}
\sqrt{\frac{R\left(R+h-R_S\right)}{R_Sh}}\,.
\label{eq:Schwresttime}
\eeq
The traveling times $\tau_R$ and $\tau$ measured by twin A's and twin B's clock, respectively, may now
be compared. Figure~\ref{fig:Schwarztimeratios} shows the ratio $\tau_R/\tau$ as a function of $R/R_S$ for
three particular choices of the height $h$. As can be seen, twin B's clock measures a larger time interval
than twin A's clock during travel, i.e., $\tau>\tau_R$. In other words, twin A, who is stationary and has
vanishing 3-acceleration, non-vanishing 4-acceleration and does not move, is now \emph{younger} at the
reunion, in agreement with the result obtained in Sec.~\ref{sec:uniformacc}. Interestingly, these
calculations thus demonstrate that vertical motion in Schwarzschild spacetime gives exactly the opposite
result of the circular motion considered by Abramowicz and Bajtlik~\cite{Rindler}.
The notions introduced by Abramowicz and Bajtlik of "a global standard of rest" and that "the twin who moves
faster with respect to the global standard of rest is younger at the reunion", appears therefore to be
irreconcilable with our results describing vertical motion in Schwarzschild spacetime.

\section{A cosmic time effect\label{sec:rotframe}}

In this section we shall consider the twin paradox in the Schwarzschild spacetime as described in a rotating reference
frame comoving with the circularly moving twin as discussed in Section~\ref{sec:intro}.

In order to gain further insight into the problem of circular motion in
Schwarzschild spacetime, discussed in Sec.~\ref{sec:intro}, we now consider the situation from the point of
view of a rotating reference frame in which the "traveling" clock (twin B's clock) is at rest. A set of coordinates
$(t',r',\theta',\phi')$, comoving with the rotating reference frame, is given by the transformation
\beq
t'=t\,,\ \ r'=r\,,\ \ \theta'=\theta\,,\ \ \phi'=\phi-\omega t\,.
\label{eq:Schwarztransforms}
\eeq
Here $\omega>0$ represents the angular velocity of the reference frame with respect to the cosmic matter
(see Sec.~\ref{sec:Mach}). For simplicity we assume that the two twins perform orbital motion at a constant
radius $r'=r_c$ in the equatorial plane for which $\theta'=\pi/2$. Then, substituting the non-zero differentials
$dt'$ and $d\phi'$ in Eq.~(\ref{eq:Schwmetric}), the line element in the rotating reference frame may be written
\beq
ds^2=-\left(1-\frac{R_S}{r_c}-\frac{r_c^2\omega^2}{c^2}\right)c^2dt'^2
+ r_c^2d\phi'^2 + 2r_c^2\omega d\phi'dt'\,.
\label{eq:rotSchwarzmetric}
\eeq
To calculate the time intervals shown by the twins' clocks at reunion, we first rewrite the line element
above as follows:
\beq
ds^2=-\left(1-\frac{R_S}{r_c}-\frac{r_c^2\omega^2}{c^2}\right.
- \left. \frac{r_c^2}{c^2}\Omega^2
- \frac{2r_c^2\omega}{c^2}\Omega\right)c^2dt'^2\,,
\label{eq:Schwarzmetrewritten}
\eeq
where $\Omega=d\phi'/dt'$ is the angular velocity of an object in the rotating reference frame in which twin B stays at
rest. From Eq.~(\ref{eq:Schwarzmetrewritten}) it is then seen that the proper time
interval measured by twin B's clock at reunion becomes
\beq
\tau_B = \left(1-\frac{R_S}{r_c}-\frac{r_c^2\omega^2}{c^2}\right)^{1/2}\Delta t'\,,
\label{eq:rottwinBtime}
\eeq
where $\Delta t'$ is the coordinate time between the twins' departure and reunion.
The last two terms entering Eq.~(\ref{eq:rottwinBtime}) both represent well-known gravitational time dilation
effects, i.e., time dilation effects caused by differences in height. The term $R_S/r_c$ represents the
gravitational time dilation a static observer outside a gravitating body would experience. The term $r_c\omega^2/c^2$ on
the other hand, represents the time dilation caused by the gravitational field experienced by an observer at
rest in a rotating reference frame.

\begin{figure}[tbp]
\begin{center}
\includegraphics[width=8.0cm]{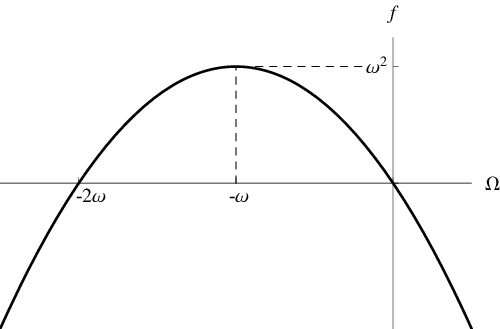}
\caption{Sketch of the function $f\left(\Omega\right)$ introduced in Eq.~(\ref{eq:rottwinAtimefunc}) for
different coordinate velocities $\Omega$.}
\label{fig:timefunc}
\end{center}
\end{figure}

Twin A, however, now moves in a circular orbit in the rotating reference frame. Consequently, the corresponding
proper time interval measured by twin A's clock becomes
\beq
\tau_A &=& \left(1-\frac{R_S}{r_c}-\frac{r_c^2\omega^2}{c^2}
+\frac{r_c^2}{c^2}f\left(\Omega\right)\right)^{1/2}\Delta t'\,,
\label{eq:rottwinAtime}
\eeq
where, for convenience, we have introduced the function
\beq
f\left(\Omega\right)= -\Omega^2 - 2\omega\Omega\,.
\label{eq:rottwinAtimefunc}
\eeq
A sketch of $f(\Omega)$ is shown in Fig.~\ref{fig:timefunc}.
Comparing Eqs.~(\ref{eq:rottwinBtime})
and (\ref{eq:rottwinAtime}) it is clear that $\tau_B$ and $\tau_A$ differ only by the two terms given by the
function $f(\Omega)$. The physical interpretation of these two terms is as follows. The term
$r_c^2\Omega^2/c^2$ simply represents another well-known relativistic effect: it is the usual
kinematical time dilation caused by relative coordinate velocities. This term is present for a moving
observer, i.e., when $|\Omega|>0$, whether or not the reference frame is rotating. The term
$-2r_c^2\omega\Omega/c^2$, however, seems to be a new kind of time dilation effect having a distinctly
different character associated only with moving clocks in rotating reference frames for which $|\omega|>0$.
This effect is similar to the Coriolis acceleration in that it is proportional
to the product of the angular velocity with respect to the cosmic matter ($\omega$) and the coordinate
angular velocity ($\Omega$). It contributes with a decrease of the rate of time if the angular velocity of
the particle in the rotating reference frame has the same direction as the angular velocity of the
reference frame relative to the stars. If these two angular velocities have opposite directions, the term
contributes with an increase of the rate of time. This new time effect will henceforth be referred to as the
\emph{cosmic time effect} for reasons that will be explained in section~\ref{sec:inertdrag}.

For the situation when $-2\omega<\Omega<0$, it is seen that $f(\Omega)$
gives a positive contribution to $\tau_A$ such that $\tau_A>\tau_B$. In particular we note that if twin A
stays at rest with respect to the cosmic matter, i.e. $\Omega=-\omega$, $\tau_A$ reaches the
maximum value $1-R_S/r_c$. For $\Omega<-2\omega$ or $\Omega>0$ on the other hand, $f(\Omega)$
becomes negative. Then the angular velocity of twin A with respect to the cosmic matter is larger
than that of twin B, and for that case $\tau_B>\tau_A$. Thus twin A's clock ages faster the less angular velocity
it has relative to the cosmic matter.

Restricting our attention to the situation $-2\omega<\Omega<0$, the following observation may now be
emphasized: even if twin B performs geodetic motion while staying at rest in the rotating reference frame whereas
twin A moves in a circular orbit at the same radius under the influence of forces, twin A ages more quickly than
twin B. In other words, the moving twin (twin A) who has a non-zero coordinate velocity and therefore
experiences a kinematical time dilation nevertheless ages faster than the twin staying at rest (twin B). The
reason for this rather peculiar result is that the moving twin also experiences the
cosmic time effect which counterbalances the kinematical time dilation. Thus we find that twin A, who has a
non-vanishing 4-acceleration, is older than his non-accelerated brother in this case. This is in agreement
with the results found by Abramowicz and Bajtlik.

Finally, we emphasize that Eqs.~(\ref{eq:rottwinBtime}) and (\ref{eq:rottwinAtime}) remain valid even in the
absence of a nearby star, i.e., in the limit $R_S\rightarrow 0$ corresponding to flat spacetime. In the
rotating reference frame, twin A then moves in a circular orbit about a central axis in the locally empty space.
For the particular situation in which $\Omega=-\omega$, twin A stays at rest with respect to the cosmic
matter. It is then twin A who has a vanishing 4-acceleration, in contrast to the situation where the star is
present.

In order to explain the peculiar fact that, for circular motion in the Scwarzschild spacetime, the stationary
twin's clock ages faster independently of the twins' 4-accelerations, Abramowicz and Bajtlik suggested that
it is the twin who moves faster with respect to the global standard of rest that is younger at the reunion.
Yet, in our universe it seems to be a basic observational fact that, at each point in spacetime, all
local inertial frames are equivalent for the performance of all physical experiments. From this it
follows that the concept of velocity is a description of \emph{relative} motion, and not a description
of motion with respect to an absolute standard of rest. This leads us to propose an alternative explanation
to the preceding results, as discussed in more detail in the following sections.

\section{A new interpretation of the Hafele-Keating experiment\label{hafelekeating}}

The version of the twin paradox with clocks in circular orbits around the Earth was realized experimentally
by J. C. Hafele and R. E. Keating in 1972~\cite{HafeleKeatingpred,HafeleKeatingobs}. Particular focus was given
to the 'East-West effect', i.e., that the travel time measured by a clock during circumnavigation of the Earth
depends both on the direction of the circumnavigation and on the Earth's rotational speed.

Hafele deduced the proper time shown by the clocks by employing a non-rotating reference frame in which the
Earth rotates~\cite{Hafele}. In the Galilean approximation the velocity of the clocks in this frame is
\beq
u = \left(R+h\right)\omega\pm v\,,
\label{eq:Hafelevel}
\eeq
where $R$ is the radius of the Earth, $h$ is the height of the orbit, $\omega$ the angular velocity of the
diurnal rotation of the Earth, and $v$ is the velocity of the airplane with plus for travelling eastwards
and minus westwards. The East-West effect then comes from the usual kinematical time dilation factor
$\sqrt{1-u^2/c^2}$.

The East-West travel time difference was interpreted as a Sagnac effect by R. Schlegel~\cite{Schlegel}, and
this interpretation was taken over by N. Ashby in his description of relativistic time effects in the satellite
clocks of the GPS-system~\cite{Ashby}. But the Sagnac experiment is concerned with light travelling along a
closed path around the axis on a rotating disk. Light travelling opposite ways is made to interfere at the
end of the paths. The change of the interference pattern with angular velocity is a measure of the difference
of the travelling times for light travelling in opposite directions. In the non-rotating laboratory frame this
time difference is due to the motion of the apparatus because of the rotation of the disk, while the light moves
around the disk. In the rotating rest frame of the apparatus the time difference is due to the different
velocities of light as measured by clocks on the disk~\cite{Gronlect}.

However in the present case there are clocks that travel, not light. Hence the 'East-West effect' should be
interpreted as an effect involving clocks only, not travelling light. The 'East-West effect' is most naturally
interpreted as a cosmic time effect which appears for moving clocks in a reference frame that rotates relative
to the 'starry sky'. The mathematical expression of the 'East-West effect', interpreted as a cosmic time effect,
follows directly from Eq.~(\ref{eq:Schwarzmetrewritten}), giving the difference in travel time to lowest order,
of a clock moving eastwards and one travelling westwards
\beq
\Delta\tau = \left(2R^2\omega\Omega/c^2\right)\tau\,,
\label{eq:difftimeeastwest}
\eeq
where $\tau$ is the travelling time, $\tau=2\pi/\Omega$. Hence the East-West time difference may be written
\beq
\Delta\tau = 4\omega A/c^2\,,
\label{eq:difftimeeastwestrew}
\eeq
where $A=\pi R^2$ is the area enclosed by the paths of the clocks. The East-West time difference is independent
of the velocity of the clocks, depending only upon the angular velocity of the reference frame and the area
enclosed by the paths of the clocks.

This effect acts so as to make the clock that has the smallest angular velocity relative to the 'starry sky'
age fastest. This means that the clock that travels westwards, i.e. in the opposite direction to the rotation
of the Earth, ages faster and will show a greater travelling time than the one that travels eastwards. The
time difference predicted by Eq.~(\ref{eq:difftimeeastwest}) with clocks moving circularly around the Earth in
the equatorial plane, at a height which is negligible compared to the radius of the Earth, is
$\Delta\tau=415$ ns. The clock moving westwards should show a travelling time \mbox{207,5 ns} longer than that shown
on a reference clock at rest at the point of departure and arrival on the Earth, and a clock moving eastwards
a \mbox{207,5 ns} shorter travelling time. The existence of this effect was confirmed by the measurements (using
commercial airplanes that did not move along circular paths in the equatorial plane. Hence, the predicted
time differences, that also had to take account of the gravitational time dilation, was made by numerical
calculation along the actual paths).

\section{Ageing in the Kerr spacetime\label{sec:Kerrspacetime}}

The twin paradox in the Kerr metric has been analyzed previously by F.~L. Markley~\cite{Markley}. He
derived the difference in the rate of time measured by two clocks moving freely in opposite directions
along circular paths around the symmetry axis of the Kerr space.

As a final example we shall in the present article consider circular motion in an axially symmetric, stationary
space which is asymptotic Minkowskian far away from $r=0$. Along the circular path the line element can then be
written
\beq
ds^2 = g_{tt}dt^2 + 2g_{t\phi}dtd\phi + g_{\phi\phi}d\phi^2\,.
\label{eq:circKerrmetric}
\eeq
Hence the proper time $\tau$ of a twin with coordinate angular velocity $\Omega=d\phi/dt$ is given by
\beq
d\tau = \left(-g_{tt}-2g_{t\phi}\Omega-g_{\phi\phi}\Omega^2\right)^{1/2}dt\,,
\label{eq:propertimeKerr}
\eeq
where $t$ is the coordinate time which is everywhere equal to the proper time measured on clocks at rest
in the asymptotic Minkowski metric.

A zero angular momentum observer (ZAMO) is defined as an observer with vanishing momentum conjugate
to the angular coordinate $\phi$, $p_{\phi}=0$, and has a coordinate angular velocity~\cite{Moller}
\beq
\Omega_Z = -\frac{g_{t\phi}}{g_{\phi\phi}}\,.
\label{eq:zamoangmomkerr}
\eeq
Hence the proper time of a ZAMO is
\beq
d\tau = \left(-g_{tt}+g_{\phi\phi}\Omega_Z^2\right)^{1/2}dt\,.
\label{eq:propertimezamo}
\eeq

Let us find the angular velocity of the circularly moving twin who ages fastest.  Naively one might think
that it is the twin at rest, since the velocity dependent time dilation tends to slow down the ageing.
Putting the derivative of the function
\beq
F\left(\Omega\right) = -g_{tt}-2g_{t\phi}\Omega-g_{\phi\phi}\Omega^2
\label{eq:funcfastestaging}
\eeq
equal to zero, one finds, however, that the ZAMO ages fastest. This is in agreement with a result of
Abramowicz and Bajtlik~\cite{AbramowiczBajtlik}. In the case of the Kerr metric the ZAMO angular velocity
is
\beq
\Omega_Z = \frac{2mac}{r^3+ra^2+2ma^2}\,,
\label{eq:zamoangvel}
\eeq
where $m=GM/c^2$ is the gravitational length of the central body with mass $M$ and $a=J/Mc$ where $J$ is its
angular momentum (note that $a$ has dimension length). Markley~\cite{Markley} found that the angular
velocities of the freely moving clocks is
\beq
\Omega=\pm\frac{\omega_1}{1\pm a\omega_1/c}\,,\,\,\,\,\,\,\omega_1=c\sqrt{\frac{m}{r^3}}\,.
\label{eq:angvelfreelymov}
\eeq
Hence there are no ZAMOs moving freely along circular paths. In the Kerr spacetime the clocks with
a constant value of $r$ that age fastest must be acted upon by a non-gravitational force.

For the Kerr metric in Boyer-Lindquist coordinates the ZAMO-angular velocity vanishes in the asymptotic
Minkowski region, and increases with decreasing radius. This is an expression of the inertial dragging
caused by the central mass in the Kerr spacetime.

\section{Inertial dragging inside a rotating shell of matter\label{sec:inertdrag}}

We have found that of all twins moving circularly around the symmetry axis of the Kerr space, the
ZAMO-twin ages fastest. The Kerr metric describes empty space outside a rotating body. There is Minkowski
spacetime in the asymptotic region far away from the body.

As seen from Eq.~(\ref{eq:zamoangvel}) ZAMO-twins at different radii have an angular velocity that vanishes in
this far away region. But what does this mean? The far away region is empty. Hence the twin with vanishing velocity
in this region must be selected in an ad hoc way.

We shall introduce an extended model of Minkowski spacetime in the next section in order to remove this
non-satisfactory character of flat spacetime. As a preparation for this we need to discuss the phenomenon of
\emph{perfect dragging} of inertial frames. By this is meant that inertial frames are dragged by a rotating or
accelerating mass distribution so that they do not rotate or accelerate relative to it.

Inertial dragging inside a rotating shell of matter was described by H. Thirring already in 1918~\cite{Thirring}.
He calculated the angular velocity, $\Omega$, of inertial frames inside a slowly rotating shell with mass $m$ in
the weak field approximation, and found that in the equation of motion of a free particle near the center of the
shell there appeared a Coriolis acceleration term corresponding to that in a frame rotating with angular velocity
\beq
\Omega = \frac{8r_s}{3r_0}\omega\,,
\label{eq:weakfomega}
\eeq
where $r_s=Gm/2c^2$ is the Schwarzschild radius of the shell and $r_0$ is its radius, both in isotropic
coordinates where the coordinate velocity of light is the same in all directions~\cite{GronHervik}.

This calculation did not, however, remove the difficulty with the asymptotically empty Minkowski spacetime.
Both the angular velocity of the shell and that of the inertial frames inside it, are defined with respect to a
system that is non-rotating in the far away region. There is nothing that determines this system. The absolute
character of rotational motion associated with the asymptotically empty Minkowski spacetime, has appeared.

However in 1966 D. R. Brill and J. M. Cohen~\cite{BrillCohenrotmass} presented a calculation of the dragging
angular velocity inside a massive shell beyond the weak field approximation. Their result is valid for
arbitrarily strong gravitational fields although they restricted the calculation to slow rotation. This
article represents a decisive step toward a resolution of the problem associated with the globally empty
Minkowski spacetime. They showed that to first order in $\omega$ spacetime inside the shell is flat. They then
deduced that the angular velocity of the inertial frames inside the shell, as expressed in isotropic
coordinates, is
\beq
\Omega = \frac{4r_s\left(2r_0-r_s\right)}{\left(3r_0-r_s\right)\left(r_0+r_s\right)}\omega\,.
\label{eq:BrillCohenangvel}
\eeq
In the case of a weak gravitational field, i.e. if the radius of the shell is much larger than its Schwarzschild
radius, $r_0\gg r_s$, this expression reduces to Eq.~(\ref{eq:weakfomega}).

On the other hand, if the shell has a radius equal to its Schwarzschild radius, $r_0=r_s$,
Eq.~(\ref{eq:BrillCohenangvel}) gives $\Omega=\omega$, i.e. perfect dragging. Brill and Cohen notes the
Machian character of this result defining the contents of "Machian" in the following way: "For mass shells
comprising more nearly all the matter in the universe than those treated by Thirring, Mach's principle
suggests that the inertial properties of space inside the shell no longer depend on the inertial frames at
infinity, but are completely determined by the shell itself." Having deduced the result above, they write
that in the limit that the radius of the shell approaches its Schwarzschild radius the rotation rate of the
inertial frames inside the shell approaches the rotation rate of the shell. In other words, in this limit
\emph{the inertial properties of space inside the shell no longer depend on the inertial frames at infinity, but
are completely determined by the shell itself}.

They further write: "A shell of matter of radius equal to its Schwarzschild radius has often been taken as an
idealized cosmological model of our universe. Our result shows that in such a model there cannot be a rotation
of the local inertial frame in the center relative to the large masses in the universe. In this sense our
result explains why the "fixed stars" are indeed fixed in our inertial frame, and in this sense the result
is consistent with Mach's principle."

In Ref.~\cite{CohenBrillgenrel} Cohen and Brill found the rotation of inertial frames induced by an uncompressible
fluid sphere and by concentric mass shells. The results were still qualitatively in agreement with the
Machian character of rotational motion. The result of Brill and Cohen were generalized to second order in $\omega$
by H. Pfister and H. K. Braun~\cite{PfisterBraun}. S. M. Lewis~\cite{Lewis} modified the deduction by placing
the shell in a FRW-universe model, and arrived at similar results. These results were extended by
C. Klein~\cite{Klein} who introduced a shell with mass equal to that cut out of the FRW-universe in the region
inside the shell.

Inertial dragging has also been discussed by several authors in the book \emph{Mach's Principle} [27]. It was pointed
out that there remain some problems concerning the relationship between Mach's principle and inertial dragging.
Pfister [28] wrote in 1995 that whether there exists a solution of Einstein's field equations with flat
spacetime and correct expressions for the centrifugal- and Coriolis acceleration inside a rotating shell of
matter, was still not known.

However, permitting singular shells such solutions certainly exist. In this connection one should note that
there is Kerr spacetime outside a rotating shell of matter, and not Schwarzschild spacetime. So the question
is whether there exists a solution of Einstein's field equations describing a rotating shell with acceptable
physical properties having Kerr spacetime outside the shell and Minkowski spacetime inside it. Furthermore, the
inertial dragging outside the shell must be so that there is perfect dragging at the shell. Continuity will then
secure perfect dragging inside the shell and correct inertial properties in this region.

In 1981 C. A. Lopez~\cite{Lopez} found a source of the Kerr spacetime. A few years later {\O}. Gr\o n~\cite{Gronsource}
gave a simple deduction of this source and discussed its physical properties, but the phenomenon of inertial
dragging was not considered. The source may be described as the surface of an oblate ellipsoid consisting of a
domain wall in which "bubbles" with negative energy rotate along the wall. The shell rotates rigidly with an
angular velocity
\beq
\omega = \frac{ac}{a^2+r_0^2}\,.
\label{eq:angvelrotshell}
\eeq
The radius of the exterior horizon in the Kerr space is
\beq
r_+ = m+\sqrt{m^2-a^2}\,
\label{eq:radexthorizon}
\eeq
and fulfills the relationship
\beq
r_+^2+a^2=2mr_+\,.
\label{eq:radrel}
\eeq
Hence, if the radius of the shell is equal to the horizon radius, $r_0=r_+$, the ZAMO-angular velocity just
outside the shell is equal to the angular velocity of the shell,
\beq
\Omega_Z\left(r_+\right) = \omega\left(r_+\right)=\frac{ac}{2mr_+}\,.
\label{eq:zamooutside}
\eeq
Note that the angular velocity of the shell is smaller the larger radius it has. The velocity of a particle
following the shell is
\beq
v = r_+\Omega_Z\left(r_+\right)=\frac{ac}{2m}\,.
\label{eq:velpartshell}
\eeq
An extreme Kerr metric has $a=m$ giving $v=c/2$. Hence the velocity is always less than (half) that of light.

Demanding continuity of the ZAMO-angular velocity it follows that the inertial frames in the Minkowski
spacetime in the interior region are co-moving with the shell. Hence there is perfect dragging of the inertial
frames inside the shell. This also means that the velocity of Eq.~(\ref{eq:velpartshell}) is not meaningful
for observers inside the shell.

The significance of this result in relation to the twin paradox is that perfect inertial dragging is necessary
in order that relativity of rotation shall be contained in the general theory of relativity, and the general
principle of relativity is needed for the formulation, i.e. for the existence of the twin paradox. If the
"travelling" twin is not allowed to consider himself as at rest, there is no twin paradox.

A conceptual difficulty discussed by J. Frauendiener~\cite{Frauendiener} should be mentioned. He writes:
"The magnitude of the dragging effect is measured by the dragging coefficient, i.e., as the ratio of the
angular velocities of the mass shell and the inertial frames in the interior. Both angular velocities are
measured by an observer at infinity. It is this reference to the preferred Minkowski frame at infinity that
has often been criticized as being "anti-Machian", as "reintroducing absolute space through the back door".

Frauendiener then goes on and introduces a different definition for the magnitude of the dragging effect which
does not refer to the infinitely distant observer. However, perfect dragging solves this problem more directly.
The observers inside the shell need no "infinitely distant observer", because with perfect dragging the rotation
is measured relatively to the shell since the non-rotating inertial frames are at rest relative to the shell.

\section{An extended model of Minkowski spacetime\label{sec:Mach}}

In our treatment of the twin paradox in the previous sections involving flat spacetime or asymptotically
flat spacetime, uniformly accelerated and rotational motion have,
without any justification, been invoked as motions with an absolute character. But the concepts of absolute
acceleration and rotation are unsatisfactory because they imply the existence of a physical space having
inertial properties which are completely independent of the physical matter contained within it. Furthermore,
a theory based on the concept of absolute space appears to be incomplete in important respects because it
cannot explain the seemingly remarkable coincidence that inertial frames are nonrotating with respect to the
fixed stars, i.e., that the swinging plane of a Foucault pendulum rotates together with 'the starry sky'.

This has been emphasized in a very clear way by J. Overduin~\cite{Overduin}. He writes that one can perform
a simple experiment on a clear night: pirouette around while looking up at the stars. You will notice two
things: one, that the stars seem to spin around in the sky, and two, that your arms are pulled upwards by
centrifugal force.

Overduin then asks whether these phenomena are connected in some way. According to Newton's dynamics
and theory of gravity they are not. This is also the case in the special theory of relativity. It is a
coincidence that stars are at rest in the same frame as the one in which we experience no centrifugal
force. Why should these two reference frames coincide? Neither Newton's theory nor the special theory
of relativity are able to answer this question.

Then Overduin notes that there are strong indications that our local "compass of inertia" has no choice
but to be aligned with the rest of the universe -- the two are linked by the frame dragging effect.
Calculations~\cite{BrillCohenrotmass,CohenBrillgenrel,Schmidframe,Schmidcosmolog} show that general
relativistic frame dragging goes over to "perfect dragging" when the dimensions of the rotating mass
become cosmological. In this limit the distribution of the matter in the universe appears sufficient
to define the inertial reference frames of observers within it.

It is usually said that the Earth bulges because it rotates. Would the Earth bulge if it were standing
still and the universe were rotating around it? Newton and special relativity say "no". But due to the
inertial dragging effect Einstein would have had the answer "yes". In this respect general relativity
is indeed more relativistic than its predecessors: it does not matter whether we choose to regard the
Earth as rotating and the heavens fixed, or the other way around. The two situations are now dynamically,
as well as kinematically, equivalent.

On the other hand, in a universe which is completely empty except for the two twins, only relative rotation
and acceleration can be meaningfully defined in a relativistic theory where there is no absolute motion.
Then a contradiction appears to arise as is illustrated by the following example.

Consider rotational motion of two twins A and B alone in an otherwise empty universe. Assume that A considers
himself as at rest at a certain distance $r$ from the central axis while B performs circular motion at the
same radius. Then, according to the kinematical laws of special relativity, A would predict that B is younger
than himself when they meet again. Yet, in the universe which is empty except for the twins, B may with equal
right think of himself as at rest and A as performing circular motion. Thus, also according to the kinematical
laws of special relativity, B would predict that A is younger than himself when they meet again. The reason for
this apparent paradox is that, in an empty universe, both twins may justifiably consider themselves as at rest
unless one introduces the concept of absolute space to break the symmetry. However, as noted by
Einstein~\cite{Einsteingenrel}, the introduction of an absolute space is merely a \emph{factitious} cause, and
not a thing that can be physically observed or measured. The idea of an absolute space should thus be rejected since it
does not provide a physically measurable \emph{mechanism} which determines why one of the twins becomes younger than
the other.

Hence, there seems to be a problem with the concept of a globally empty Minkowski spacetime: In order to break the
symmetry and distinguish between the inertial and non-inertial motion of the two twins, one must introduce the
4-vector quantities called acceleration and vorticity. But how do we assign an acceleration-vector or vorticity-vector
to a point of the empty absolute space if this space is not physically observable? In other words, how do we connect
the model of a globally empty Minkowski spacetime to our physical universe?

Our answer to these questions is that one has to take into account the gravitational action of the cosmic masses.
We propose that the concept of a globally empty, flat spacetime should be replaced by an extended model of the Minkowski
spacetime maintaining the flatness of the spacetime while it simultaneously takes into account
the gravitational effect of accelerated cosmic masses. In this extended model, introduced below, accelerated and
rotational motion are then to be perceived as truly relative concepts. The construction of such a model is possible
due to the phenomenon of \emph{perfect inertial dragging}, and the model will then be consistent with experimental
observations. Also, perfect inertial dragging provides a physical mechanism explaining the asymmetry between the two
twins' motion. Thus the notion of an unobservable absolute space becomes superfluous.

The interested reader may now object to this and point out that the special theory of relativity presupposes the
existence of inertial reference frames. The inertial reference frames are operationally determined by the local
spin axes of gyroscopes and, as long as one makes sure that the local axes of the chosen reference frame are
not rotating or accelerating with respect to the gyroscopes' axes, free particles will travel along straight
lines in this reference frame. Hence, it is tempting to to conclude that inertial reference frames are operationally
well-defined without any reference whatsoever to the cosmic masses.

However, we note that it is a well established
experimental fact that spin axes of gyroscopes precess with the ZAMO angular velocity. In the asymptotic
Minkowski region this means that they do not precess with respect to the distant masses. As noted by
C. Schmid~\cite{Schmidframe}, no explanation for this observational fact is given in classical mechanics,
special relativity or general relativity for \emph{isolated} systems in asymptotic Minkowski space. This suggests that
rotation should be considered as a description of motion relative to the average motion of all the matter in
the universe. Furthermore, since the density approaches zero as $t\rightarrow\infty$ in a radiation and
dust dominated FRW universe, spacetime in such a universe approaches the Minkowski spacetime~\cite{CoopersFarao}.
This may be verified by calculating Kretschmann's curvature scalar~\cite{GronHervik} for the FRW metric
given by Eq.~(\ref{eq:FRWmetric}). In the limit as $a(t)\rightarrow\infty$ Kretschmann's curvature scalar
approaches zero which shows that the FRW spacetime converges to flat spacetime. Hence, the inertial frames
even in this asymptotic Minkowski spacetime do not rotate relative to the average motion of all the matter
in the universe.

The validity of this Machian interpretation of rotational motion in our universe has recently been demonstrated
by C. Schmid~\cite{Schmidframe,Schmidcosmolog}. By introducing a rotational perturbation he has shown that perfect
dragging is a property possessed by our universe, i.e., that the ZAMO angular velocity in a rotating perturbation
of a FRW universe is equal to the average angular velocity of the cosmic mass distribution
(the angular velocity of the rotating perturbation). Perfect inertial dragging explains why
the swinging plane of a Foucault pendulum rotates together with "the starry sky". In Newtonian
gravity, where there is no inertial dragging, this is only a strange coincidence.

In this context, it should be noted that one makes an implicit assumption when one introduces asymptotic
Minkowski spacetime in the solutions of Einstein's equations for the geometry outside a localized mass
distribution, namely that the ZAMOs are non-rotating relative to cosmic masses in this region.
Hence, it is the interactions between \emph{all} the masses in the universe, \emph{including nearby rotating stars
and planets}, which determine the local ZAMOs.

Operatively a local inertial frame of the special class in which the co-moving observers are ZAMOs, may be
determined by letting an observer with zero angular velocity relative to the average motion of the cosmic mass,
fall freely from a region far from any localized mass.

The considerations above suggest that Minkowski spacetime should not be thought of as globally empty. We propose
a generalized model of Minkowski space, i.e. of globally flat spacetime or the flat region of asymptotically flat
spacetimes, where the space is completed by a far away cosmic massive shell with radius equal to its Schwarzschild
radius, representing the cosmic mass. Inside such a shell there is approximately flat spacetime and perfect
dragging~\cite{BrillCohenrotmass,CohenBrillgenrel}, and section~\ref{sec:inertdrag} above. Close to
the shell the ZAMOs have no rotational motion relative to the shell. But far inside the shell the ZAMOs will in
general rotate relative to the shell due to localized rotating mass distributions.

This model of Minkowski spacetime has a Machian character. With this model the notion that rotational and
accelerated motion can only have a relative significance may be valid.

This model is also relevant in connection with a point made several years ago by Christian M\o ller~\cite{Moller}.
He wrote that when one solves Einstein's field equations in a rotating reference frame it is necessary to take account
of the far away cosmic masses. Earlier there was an exception for globally or asymptotic Minkowski spacetime,
since it contained no far away masses. In the extended model the Minkowski spacetime must be treated in the
same way as any other spacetime.

The significance of the extended model of the Minkowski spacetime for the twin paradox is the following.
The special theory of relativity is based upon the empirical fact that there exists privileged reference frames in
our universe called inertial frames, and the fundamental principles of the special theory are only valid for inertial
observers. This ensures that the physical situation usually termed the twin paradox does not imply any inconsistency
in the special theory of relativity, as was once claimed by H. Dingle~\cite{Dingle}. However, within the special theory
of relativity, we have no possibility of identifying a physical mechanism that is the cause of the asymmetry
between inertial and non-inertial observers. Invoking an absolute space is of no help, since it is nothing
that can be physically observed. The same problem is encountered with the asymptotically empty Minkowski spacetimes
used in the general theory of relativity. Our extended model of the Minkowski spacetime seem to resolve these issues.
In this model, the physical mechanism that causes the asymmetry between the two twins' motion is perfect inertial
dragging. This phenomenon requires taking into account the cosmic masses. Moreover, in the extended model
acceleration and rotation are treated as truly relativistic concepts in a similar manner as the relativistic
concepts of position and velocity. Thus, even when one considers motion of arbitrary kind, the introduction of
an unobservable absolute space is superfluous in this model.

\section{Conclusion}

Our examples have shown that spacetime curvature or some absolute standard of rest is not a decisive factor in
determining which of two twins that move away from each other and then unites again, is oldest at the reunion.
Due to the Lorentz invariance of physical phenomena there does not exist any absolute standard of rest. However,
in a universe with perfect dragging of inertial frames there exists a standard of non-acceleration and
non-rotation. This is, however, a standard defined relative to the average motion of the cosmic matter. Also
it is apparent only asymptotically far away from localized mass distributions. In such far away regions the
inertial frames are freely moving ZAMO frames modulo a velocity.

In their Schwarzschild example Abramowicz and Bajtlik formulated an interesting and seemingly paradoxical limit:
\beq
\lim_{M\rightarrow 0} \left( \begin{array}{c}
\mbox{the higher velocity twin is younger}\\
\mbox{acceleration is not important} \end{array} \right) \nonumber \\
= \left( \begin{array}{c}
\mbox{the accelerated twin is younger}\\
\mbox{velocity is not important} \end{array} \right)\,.
\label{eq:Abramowiczlimit}
\eeq
The limit was formulated in the context of circular motion, and hence vertical motion in a gravitational field
was neglected. This limit can be understood in light of the extended model of the Minkowski spacetime. The
asymptotic flatness of the Schwarzschild, or for that matter the Kerr spacetime, means that in the case of
circular motion the twin who moves most slowly relative to the local ZAMO-field ages fastest. Hence the
limit~(\ref{eq:Abramowiczlimit}) should be replaced by
\beq
\lim_{M\rightarrow 0} \left( \begin{array}{c}
\mbox{the twin who moves fastest relative}\\
\mbox{to the local ZAMO-field is younger} \end{array} \right) \nonumber \\
= \left( \begin{array}{c}
\mbox{the twin who moves fastest relative}\\
\mbox{to the cosmic matter is younger} \end{array} \right)\,.
\label{eq:modifiedAbramowiczlim}
\eeq

However, taking into account vertical motion in a gravitational field neither of these limits are generally valid.
Our examples indicate that there is no simple answer to the question "which twin is older at the reunion?". We
have shown that in order to give a general answer to this question one has to take into account both the
kinematical and gravitational time effects, and also a third  one which to our knowledge has not been mentioned
earlier: a cosmic time effect, which only appears for clocks moving in a rotating
reference frame, i.e., when $\Omega$ in Eq.~(\ref{eq:Schwarzmetrewritten}) is non-vanishing.

Our solution to the twin paradox has a Machian character. There is no absolute standard of rest because even
perfect dragging does not fix the velocity of the inertial frames. In other words perfect inertial dragging
does not destroy the Lorentz invariance of inertial frames. It fixes, however, the angular velocity of the
ZAMO-field in the asymptotic Minkowski region of a limited mass distribution.

The ZAMO-field further determines
which twin will be youngest at reunion in the case of circular motion. Consider two twins moving along circular
orbits with equal radii around a rotating body. The twin who rotates faster relative
to the ZAMOs is younger at the reunion. In general such a twin will not move freely. In the limit that the spin
of the rotating body vanishes, the angular velocity of a twin who is also a ZAMO, vanishes, and the twin will
stay at rest. If the body is spherical, this limit represents the Schwarzschild case where the ZAMO-twin is at rest
and ages faster than the one moving freely along a circular path. However, as was shown above, the twin moving
freely first upwards and then falling down again, ages faster than even the twin at rest.

Hence, moving freely is not decisive for ageing fast. It is important to be high up in a gravitational field.
Move a twin to a great height, let him stay there for a long time, and then let him come down to his brother.
Then he will be older than his brother. Clocks on a GPS-satellite, for example, age faster than clocks at rest
on the surface of the Earth, although the satellites move in the rest frame of the Earth.

When the twin paradox is described in the rotating rest frame of a twin, one has to take into account a new time
effect which involves the angular velocity of the other twin in this frame. The physics of this is that the new
effect secures that, as analyzed in any frame, the twin at rest relative to the ZAMO-field will age fastest.
Hence there is a cosmic connection here since the asymptotic ZAMO-field is determined by the cosmic matter and
energy. This is the reason we have talked about the twin paradox in a cosmological context in the heading.

The kinematic, gravitational and cosmic time effects act together and determine which twin is the
youngest one at the reunion. Curvature is not decisive. Neither some cosmic standard of rest. That does
not exist. As was shown above, circular and vertical motion in the Schwarzschild spacetime give opposite
results. With circular motion the twin A who is at rest relative to the star is oldest, but with vertical
motion the resting twin is youngest. From the point of view of A this difference is due to the fact that
the gravitational time dilation is of no significance with circular motion, but it is dominant with
vertical motion. From the point of view of the twin B moving along a circular orbit, the situation
is rather strange. B observes that A moves along a circular orbit in the same height as himself. There
is no gravitational time dilation, only a kinematical one which apparently should make A younger
than himself. Yet he finds that A is older than himself at the reunion. In order to explain this he
has to invoke the relativistic time effect which we have called the cosmic time effect. This says
that the cosmic mass acts upon the rate of time of a clock in a rotating reference frame so that
it ages faster the less angular velocity it has relative to the cosmic matter.

However, since the cosmic time dilation acts together with the kinematic and gravitational time dilation,
the clock that ages fastest is in general neither inertial nor the one with least velocity or the one
highest up in a gravitational field. It is the sum of all the effects that matters.

Finally it should be noted that a real twin paradox would arise if the general principle
of relativity was valid in a globally empty Minkowski spacetime. This paradox is resolved by the
extended model of the Minkowski spacetime.


\begin{thebibliography}{99}
\bibitem{Marder} L. Marder, \emph{Time and the Space-Traveller} (George Allen \& Unwin, London, 1971).
\bibitem{EriksenGron} E. Eriksen and \O. Gr\o n, "Relativistic dynamics in uniformly accelerated
reference frames with application to the clock paradox", Eur. J. Phys. \textbf{11}, 39-44 (1990).
\bibitem{DebsRedhead} T. A. Debs and M. G. L. Redhead, "The twin 'paradox' and the conventionality
of simultaneity", Am. J. Phys. \textbf{64}, 384-392 (1995).
\bibitem{Nicolic} H. Nicoli\'{c}, "The role of acceleration and locality in the twin paradox",
Found. Phys. Lett. \textbf{13}, 595-601 (2000).
\bibitem{Iorio} L. Iorio, "An analytical treatment of the Clock Paradox in the framework of the
special and general theories of relativity", Found. Phys. Lett. \textbf{18}, 1-19 (2006).
\bibitem{GronHervik} \O. Gr\o n and S. Hervik, \emph{Einstein's General Theory of Relativity}. (Springer, New York, 2007).
\bibitem{LichteneggerIorio} H. I. M. Lichtenegger and L. Iorio, "The twin paradox and Mach's principle",
ArXiv: 0910.1929 (2009).
\bibitem{AbramowiczBajtlikphoton} M. A. Abramowicz, S. Bajtlik and W. Klu\'{z}niak, "The twin paradox on
the photon sphere", Phys. Rev. A \textbf{75}, 044101-1-2 (2007).
\bibitem{AbramowiczBajtlik} M. A. Abramowicz and S. Bajtlik, "Adding to the paradox: the accelerated
twin is older", ArXiv: 0905.2428 (2009).
\bibitem{Einstein} A. Einstein, "Dialog \"{u}ber Einw\"{a}nde gegen die Relativit\"{a}tstheorie", Die
Naturwissenschaften \textbf{6}, 697- 702 (1918).
\bibitem{GronEriksen} \O. Gr\o n and E. Eriksen, "Translational Inertial Dragging", Gen.Rel.Grav. \textbf{21},
105-124 (1989).
\bibitem{HolsteinSwift} B. H. Holstein and A. R. Swift, "The Relativity Twins in Free Fall",
Am. J. Phys. \textbf{40}, 746-750 (1972).
\bibitem{Gron} \O. G. Gr\o n, "The principle of relativity and inertial dragging", Am. J. Phys. \textbf{77},
373-380 (2009).
\bibitem{Boughn} S. P. Boughn, "The case of the identically accelerated twins", Am. J. Phys. \textbf{57},
791-793 (1989).
\bibitem{Styer} D. F. Styer, "How do two moving clocks fall out of sync? A tale of trucks, threads and twins",
Am. J. Phys. \textbf{75}, 805-814, (2007).
\bibitem{GiannoniGron} C. Giannoni and \O. Gr\o n, "Rigidly connected accelerated clocks", Am. J. Phys. \textbf{47},
431-435 (1979).
\bibitem{Minguzzi} E. Minguzzi, "Differential aging from acceleration, an explicit formula",
Am. J. Phys. \textbf{73}, 876-880 (2005).
\bibitem{Rindler} The fact that clocks moving freely in the vertical direction and along a circular path
in the Schwarzschild spacetime age differently was noted several years ago by W. Rindler,
\emph{Essential Relativity}, Springer (1. edition 1969), Exercise 8.13.
\bibitem{HafeleKeatingpred} J. C. Hafele and R. E. Keating, "Around the world atomic clocks: predicted
relativistic time gains", Science \textbf{177}, 166-168 (1972).
\bibitem{HafeleKeatingobs} J. C. Hafele and R. E. Keating, "Around the world atomic clocks: observed relativistic
time gains", Science \textbf{177}, 168-170 (1972).
\bibitem{Hafele} J. C. Hafele, "Relativistic Time for Terrestrial Circumnavigation", Am. J. Phys.
\textbf{40}, 81-85 (1972).
\bibitem{Schlegel} R. Schlegel, "Comments on the Hafele-Keating Experiment"  Am. J. Phys. \textbf{420},
183-187 (1974).
\bibitem{Ashby} N. Ashby, "Relativity in the Global Positioning System", Living Reviews in Relativity, 2003-1.
\bibitem{Gronlect} \O. Gr\o n, \emph{Lecture Notes on the General Theory of Relativity}, Springer (2009).
\bibitem{Markley} F. L. Markley, "Relativity Twins in the Kerr Metric", Am. J. Phys. \textbf{41},
1246-1250 (1973).
\bibitem{Moller} C. M\o ller, \emph{The theory of Relativity} (Oxford, 1952), Chap. 8.
\bibitem{Thirring} H. Thirring, "\"{U}ber die Wirkung rotierender ferner Massen in der Einsteinschen
Gravitationstheorie", Physikalische Zeitschrift \textbf{19}, 33-39 (1918).
\bibitem{BrillCohenrotmass} D. R. Brill and J. M. Cohen, "Rotating masses and their effect on inertial frames",
Phys. Rev. \textbf{143}, 1011-1015 (1966).
\bibitem{CohenBrillgenrel} J. M. Cohen and D. R. Brill, "Further examples of "Machian" effects of rotating
bodies in general relativity", Nuovo Cimento \textbf{54}, 209-218 (1968).
\bibitem{PfisterBraun} H. Pfister and H. K. Braun, "Induction of correct centrifugal force in rotating mass shell",
Class. Quantum Grav. \textbf{2}, 909-918 (1985).
\bibitem{Lewis} S. M. Lewis, "Machian Effects in Nonasymptotically Flat Space-Times", Gen. Rel. Grav. \textbf{12},
917-924 (1980).
\bibitem{Klein} C. Klein, "Rotational perturbations and frame dragging in Friedmann universe", Class. Quantum
Grav. \textbf{10}, 1619-1631 (1993).
\bibitem{BarbourPfister} \emph{Mach's Principle}. Ed.: J. Barbour and H. Pfister. Birkhauser (1995).
\bibitem{Pfister} H. Pfister, "Dragging Effects Near Rotating Bodies and in Cosmological Models",
in Ref.~\cite{BarbourPfister}.
\bibitem{Lopez} C. A. Lopez, "Extended model of the electron in general relativity", Phys. Rev. D \textbf{30},
313-316 (1984).
\bibitem{Gronsource} \O. Gr\o n, "New derivation of Lopez's source of the Kerr-Newman field",
Phys. Rev. D \textbf{32}, 1588-1589 (1985).
\bibitem{Frauendiener} J. Frauendiener, "On the Interpretation of Dragging Effects in Rotating Mass Shells".
In Ref.~\cite{BarbourPfister}.
\bibitem{Overduin} J. Overduin, "Spacetime and Spin", Homepage of Gravity Probe B, http://einstein.stanford.edu/
SPACETIME/spacetime4.html.
\bibitem{Einsteingenrel} A. Einstein, "Die Grundlage der allgemeinen Relativit\"{a}tstheorie,
Annalen der Physik \textbf{49} (1916).
\bibitem{CoopersFarao} V. Faraoni and F. I. Cooperstock, "On the total energy of open Friedmann-Robertson-Walker
universes", Astrophys. J. \textbf{587}, 483-486 (2003).
\bibitem{Schmidframe} C. Schmid, "Cosmological gravitomagnetism and Mach's principle", Phys. Rev. D \textbf{74},
044031 (2006).
\bibitem{Schmidcosmolog} C. Schmid, "Mach's principle: Exact frame-dragging via gravitomagnetism in perturbed
Friedmann-Robertson-Walker universes with $K=(\pm 1,0)$", Phys. Rev. D \textbf{79}, 064007 (2009).
\bibitem{Dingle} H. Dingle, \emph{Science at the Crossroads}, (Martin Brian \& O'Keeffe, London, 1972).
\end{thebibliography}
\end{document}